\documentclass[aps,showpacs,showkeys]{revtex4}
\usepackage{amsmath,amsthm,amssymb,epsfig,alltt}
\usepackage{mathtools}

\begin{document}

\def\a{\alpha}
\def\b{\beta}
\def\d{{\delta}}
\def\l{\lambda}
\def\e{\epsilon}
\def\p{\partial}
\def\m{\mu}
\def\n{\nu}
\def\t{\tau}
\def\th{\theta}
\def\s{\sigma}
\def\g{\gamma}
\def\G{\Gamma}
\def\o{\omega}
\def\r{\rho}
\def\z{\zeta}
\def\D{\Delta}
\def\half{\frac{1}{2}}
\def\hatt{{\hat t}}
\def\hatx{{\hat x}}
\def\hatp{{\hat p}}
\def\hatX{{\hat X}}
\def\hatY{{\hat Y}}
\def\hatP{{\hat P}}
\def\haty{{\hat y}}
\def\whatX{{\widehat{X}}}
\def\whata{{\widehat{\alpha}}}
\def\whatb{{\widehat{\beta}}}
\def\whatV{{\widehat{V}}}
\def\hatth{{\hat \theta}}
\def\hatta{{\hat \tau}}
\def\hatrh{{\hat \rho}}
\def\hatva{{\hat \varphi}}
\def\barx{{\bar x}}
\def\bary{{\bar y}}
\def\barz{{\bar z}}
\def\baro{{\bar \omega}}
\def\barpsi{{\bar \psi}}
\def\sp{\sigma^\prime}
\def\nn{\nonumber}
\def\cb{{\cal B}}
\def\2pap{2\pi\alpha^\prime}
\def\wideA{\widehat{A}}
\def\wideF{\widehat{F}}
\def\beq{\begin{eqnarray}}
 \def\eeq{\end{eqnarray}}
 \def\4pap{4\pi\a^\prime}
 \def\xp{{x^\prime}}
 \def\sp{{\s^\prime}}
 \def\ap{{\a^\prime}}
 \def\tp{{\t^\prime}}
 \def\zp{{z^\prime}}
 \def\rp{\rho^\prime}
 \def\zetap{\zeta^\prime}
 \def\op{\omega^\prime}
 \def\xpp{x^{\prime\prime}}
 \def\xppp{x^{\prime\prime\prime}}
 \def\barxp{{\bar x}^\prime}
 \def\barxpp{{\bar x}^{\prime\prime}}
 \def\barxppp{{\bar x}^{\prime\prime\prime}}
 \def\barchi{{\bar \chi}}
 \def\baro{{\bar \omega}}
 \def\bpsi{{\bar \psi}}
 \def\barg{{\bar g}}
 \def\barz{{\bar z}}
 \def\bareta{{\bar \eta}}
 \def\ta{{\tilde \a}}
 \def\tb{{\tilde \b}}
 \def\tc{{\tilde c}}
 \def\tz{{\tilde z}}
 \def\tJ{{\tilde J}}
 \def\tpsi{\tilde{\psi}}
 \def\tal{{\tilde \alpha}}
 \def\tbe{{\tilde \beta}}
 \def\tga{{\tilde \gamma}}
 \def\tchi{{\tilde{\chi}}}
 \def\barth{{\bar \theta}}
 \def\bareta{{\bar \eta}}
 \def\barom{{\bar \omega}}
 \def\bole{{\boldsymbol \epsilon}}
 \def\bolth{{\boldsymbol \theta}}
 \def\bomega{{\boldsymbol \omega}}
 \def\bolmu{{\boldsymbol \mu}}
 \def\bola{{\boldsymbol \alpha}}
 \def\bolb{{\boldsymbol \beta}}
 \def\bolX{{\boldsymbol X}}
 \def\bolvphi{{\boldsymbol \varphi}}
 \def\bolv{{\boldsymbol v}}
 \def\mathN{{\boldsymbol n}}
 \def\bba{{\boldsymbol a}}
 \def\bbA{{\boldsymbol A}}
 \def\mathP{{\mathbb P}}
 \def\mathN{{\boldsymbol N}}
 \def\mathN{{\mathbb N}}
 \def\bbP{{\boldsymbol P}}
 \def\bbk{{\boldsymbol k}}
 
\setcounter{page}{1}
\title[]{Covariant Open String Field Theory on Multiple $Dp$-Branes
}

\author{Taejin Lee}
\affiliation{
Department of Physics, Kangwon National University, Chuncheon 24341
Korea}

\email{taejin@kangwon.ac.kr}

\begin{abstract}
We study covariant open bosonic string field theories on multiple $Dp$-branes by using the deformed cubic string field theory which is equivalent to the string field theory in the proper-time gauge. Constructing the Fock space representations of the three-string vertex and the four-string vertex on multiple $Dp$-branes, we obtain the field theoretical effective action in the zero-slope limit. On the multiple $D0$-branes, the effective action reduces to the Banks-Fishler-Shenker-Susskind (BFSS) matrix model. 
We also discuss the relation between the open string field theory on multiple $D$-instantons in the 
zero-slope limit and the Ishibashi-Kawai-Kitazawa-Tsuchiya (IKKT) matrix model. 
The covariant open string field theory on multiple $Dp$-branes would be useful to study the
non-perturbative properties of quantum field theories in $(p+1)$-dimensions in the framework of the string theory. The non-zero-slope corrections may be 
evaluated systematically by using the covariant string field theory.  

\end{abstract}


\pacs{11.15.−q, 11.25.Uv, 11.25.Sq }

\keywords{open string, $Dp$-brane, covariant string field theory, Yang-Mills gauge theory, matrix model}

\maketitle

\section{Introduction}

The string theories are defined only in critical dimensions; $10$ dimension for the super-string 
theories and $26$ dimension for the bosonic string theories. However, the quantum field theories, which describe 
open strings in low energy region can be defined in any dimension less than or equal to the critical 
dimension $d_{\text{critical}}$ if we construct the string field theories on $Dp$-branes, $ -1 \le p \le d_{\text{critical}}-1$. Thus, the string field theory provides a unique framework to explore low dimensional quantum field theories in a unified manner. The purpose of this work is twofold: First, we shall construct
covariant string field theories on $Dp$-branes of which zero-slope limits correspond to the quantum field theories in dimension lower than the critical dimension. These covariant string field theories will be useful to understand various non-perturbative features of quantum field theories, which could not have been approached by the conventional perturbation theory. Second, we wish to understand the origins of actions for the matrix models \cite{BFSS, Ishibashi1997}, which have served as important tools to study the non-perturbative effects of super-string theories and the $M$-theory \cite{Hull94,Witten95s,Duff95} within the framework of the covariant string field theory. 

The core strategy we shall adopt in the present work is the deformed cubic open string field theory \cite{Lee2016i,Lee2017d}, which is equivalent to the covariant string field theory in the proper-time gauge \cite{Lee88}. We have shown that the 
deformed cubic open string field theory if defined on the space filling $D$-brane, yield the non-Abelian Yang-Mills theory in the zero-slope limit. The main reason we adopt the deformed cubic open string field theory is that we can obtain the exact results without using the field redefinition \cite{Feng2007} or the level truncation \cite{Kostecky90,Kostecky96,Sen99,Taylor2000,Coletti03}. 
The deformed cubic string field theory may also provide a systematic means to calculate the non-zero-slope 
corrections \cite{Tseytlin86} and string scattering amplitudes \cite{JCLee2015,Lai2016,Huang2016a,Huang2016b,Lai2017}. 
In fact, deformation of the cubic interaction is not a new idea. Hua and Kaku \cite{Huakaku} has discussed deformation of the midpoint overlapping interaction of Witten's cubic string field theory into the
endpoint interaction in the context of closed string field theory. 
In recent works \cite{Lee2016i,Lee2017d} we developed the deformed cubic open string
field theory by defining the theory on space filling $D$-branes. On space filling $D$-branes, the end 
points of the string satisfy only the Neumann boundary condition, so that the light-cone string field theory 
technique \cite{Mandelstam1973,Mandelstam1974,Kaku1974a,Kaku1974b,Cremmer74,Cremmer75,GreenSW} was readily available. To deal with the open strings on the multiple $Dp$-branes, of which 
string coordinates along the directions, orthogonal to the $Dp$-brane worldvolume, satisfy the Dirichlet boundary condition, we need to extend the previous works appropriately.

The deformation procedure transforms the non-planar world sheet diagrams of the Witten's cubic open string field theory \cite{Witten1986,Witten92p} into equivalent planar diagrams of the string field theory in the proper-time gauge. 
In the present work, we shall show that the deformation procedure is also applicable to the 
open string which satisfies the Dirichlet boundary condition. 
Then by mapping the planar diagrams of the deformed cubic string field theory on multiple $Dp$-branes onto the upper half plane, we will be able to evaluate the Neumann functions of the three-string vertex and the four-string vertex for the string
on multiple $Dp$-branes. With the Neumann functions, we shall construct the Fock space representation of the string vertices and calculate the three-string and the four-string scattering amplitudes.
In the zero-slope limit the external string states are $U(N)$ matrix valued 
non-Abelian gauge fields and $(d_{\text{critical}}-p-1)$ scalar fields in $(p+1)$ dimensions.
From the three-string scattering amplitude and the four-string scattering amplitude in the zero-slope limit,
we get the correct $U(N)$ gauge invariant matrix valued scalar field theory, which describes dynamics of the multiple $Dp$-branes in the low energy region. In particular, for the multiple $D0$-branes the covariant 
open string field theory reduces to the $U(N)$ matrix quantum mechanics, which has been the main subject of the 
Banks-Fishler-Shenker-Susskind (BFSS) matrix model \cite{BFSS}. Choosing the multiple $D$-instantons may bring us an open string field theory of which action can be expressed solely in terms of matrices. In the zero-slope limit, the cubic 
string field theory on the multiple $D$-instantons is expected to reduce to the Ishibashi-Kawai-Kitazawa-Tsuchiya (IKKT) matrix model \cite{Ishibashi1997} of which action comprises only the contact quartic term of $U(N)$ matrix valued vector fields.

\section{Open String Fields on $Dp$-Branes} 

On a $Dp$-brane, the string coordinates $X^\m$, $\m = 0, 1, \dots, p$ are tangential to the $Dp$-brane 
world-volume and the string coordinates $X^i$, $i=p+1, \dots, d=d_{\text{critical}}-1$, are normal to the 
$Dp$-brane world-volume: The end points of $X^\m$, $\m = 0, 1, \dots, p$ satisfy the Neumann condition
and the end points of $X^i$, $i=p+1, \dots, d$ satisfy the Dirichlet condition
\begin{subequations}
\beq
\frac{\p X^\m}{\p \s} \Bigl\vert_{\s = 0, \, \pi} &=& 0, ~~~\text{for}~~~ \m =0, 1, \dots, p, \\
X^i \Bigl\vert_{\s = 0, \, \pi}  &=& 0, ~~~ \text{for} ~~~ i = p+1, \dots, d . 
\eeq 
\end{subequations}
In accordance with the boundary conditions,  
the string coordinates $X^I$, $I = 0, 1, \dots , d$ may be expanded in terms of the normal modes as
\begin{subequations}
\beq
X^\m (\s) &=& x^\m + \sqrt{2} \sum_{n=1} x^\m_n \cos \left(n \s\right), ~~~ \m = 0, 1, \dots, p, \label{xmodem} \\
X^i (\s) &=& \sqrt{2} \sum_{n=1} x^i_n \sin \left(n \s\right), ~~~ i = p+1, \dots, d. \label{xmodei}
\eeq
\end{subequations}
Note that the string coordinates $X^i$, $i=p+1, \dots, d$ do not contain zero modes. 

The string propagator is obtained by evaluating the path integral on a strip with the Polyakov string action 
\begin{subequations}
\beq
G[X_1; X_2] &=& \int D[h] D[X] \exp \left(iS \right), \\
S &=& - \frac{1}{4\pi \ap} \int_M d\t d\s \sqrt{-h}h^{\a\b} \frac{\p X^I}{\p \s^\a} \frac{\p X^J}{\p \s^\b} 
\eta_{IJ}, ~~~ I, J = 0, \dots, d
\eeq 
\end{subequations}
where $\s^1 = \t$, $\s^2 = \s$ and $\ap$ is the Regge slope parameter. 
We may fix the reparametrization invariance
by choosing the proper-time gauge where the proper-time on the string world sheet is defined properly \cite{Lee88},
\beq
\p_\t N_{10} = 0, ~~~ N_{1n} = 0, ~~~ N_{2n} = 0, ~~~ n\not=0,
\eeq 
where $N_{\a n}$ is the normal modes of the lapse and shift functions 
$N_\a = \sum_n N_{\a n} e^{in\s}$, $\a=1, 2$ of the two dimensional metric on the 
world sheet 
\beq
\sqrt{-h}h^{\a\b} = \frac{1}{N_1} \left(\begin{array}{rc} - 1  & N_2 \\ N_2 & (N_1)^2 -(N_2)^2 \end{array} \right).
\eeq 
Evaluating the Polyakov path integral leads us to the open string field propagator on the $Dp$-branes
\begin{subequations}
\beq
G[X_1; X_2) &=& \int_0^\infty ds \langle X_1 \vert \exp \left[-is \left(L_0 -i\e\right) \right] 
\vert X_2 \rangle \nn\\
&=& \langle X_1 \vert \frac{1}{L_0 -i\e} \vert X_2 \rangle, \\
L_0 &=& \frac{p^\m p_\m}{2} + \sum_{n=1} \half \left(p^I_n p^J_n + n^2 x^I_n x^J_n \right)\eta_{IJ} -1, 
\eeq 
\end{subequations}
where $p^I_n$, $I=0, 1, \dots, d$ are normal modes of the momentum operators $P^I$
\begin{subequations}
\beq
P^\m(\s) &=& \frac{1}{\pi} \left(p^\m + \sqrt{2}\sum_{n=1} p^\m_n \cos\left(n\s\right)\right), ~~~
\m = 0, 1, \dots, p, \label{pmodem}\\
P^i(\s) &=& \frac{\sqrt{2}}{\pi} \sum_{n=1} p^i_n \sin \left(n\s\right), ~~~ i = p+1, \dots, d. \label{pmodei}
\eeq 
\end{subequations}
(Throughout this paper, we suppress the ghost sector for the sake of simplicity.)

Because the end point of the open string is attached on one of $N$ $Dp$-branes, the open string has $N^2$ 
different quantum states and consequently, the string field $\Psi$ carries the group indices of $U(N)$
\beq
\Psi [X] &=& \frac{1}{\sqrt{2}} \Psi^0[X] + \Psi^a[X] T^a, ~~~ a =1, \dots, N^2 -1
\eeq
where $T^a$ $a=1, \dots, N^2 -1$ are generators of $SU(N)$ group. Now the string propagator on the multiple 
$Dp$-branes, carrying the group indices, may be written as 
\beq
G^{ab} [X_1, X_2] &=& i \langle T \Psi^a[X_1] \Psi^b[X_2] \rangle \nn\\
&=& i \int D[X] \Psi^a[X_1] \Psi^b[X_2] \exp \Biggl\{-i \int D[X] \, \text{tr}\, \Psi \left(L_0 + i\e \right) 
\Psi  \Biggr\} .
\eeq 
From this expression of the string propagator, the action of the string field theory follows 
\beq
{\cal S}_0 = \int D[X] \, \text{tr}\, \Psi \left(L_0 + i\e \right) \Psi .
\eeq
If we introduce the BRST ghosts, we may cast the free string field action into a BRST invariant form
\beq
{\cal S}_0 &=& \int \, \text{tr} \, \Psi * Q \Psi  
\eeq 
where $Q$ is the BRST operator. 

\section{Deformation of Cubic Open String Field Theory on Multiple $Dp$-Branes}

It is not difficult to extend the Witten's cubic open string field theory \cite{Witten1986} defined on a space filling $D$-brane
to the cubic open string field theory on the multiple $Dp$-branes. It only takes 
replacing normal mode expansions of the string coordinates $X^I$ and the momentum operators $P^I$, 
$I = 0, 1, \dots, d$ by those given as Eqs. (\ref{xmodem}, \ref{xmodei}) and Eqs. (\ref{pmodem}, \ref{pmodei}): 
\beq
{\cal S} &=& \int \text{tr} \left( \Psi * Q \Psi + \frac{2g}{3} \Psi * \Psi * \Psi \right),
\eeq
where the star product between the string field operators is defined as
\begin{subequations}
\beq
\left(\Psi_1 * \Psi_2\right) [X(\s)] &=& \int \prod_{\frac{\pi}{2} \le \s \le \pi} DX^{(1)}(\s) \prod_{0 \le \s \le \frac{\pi}{2}} DX^{(2)}(\s)  \nn\\
&& 
\prod_{\frac{\pi}{2} \le \s \le \pi} \d \left[X^{(1)}(\s) - X^{(2)}(\pi -\s) \right] \Psi_1[X^{(1)}(\s)] \Psi_2[X^{(2)}(\s)], \\
X(\s) &=& \left\{ \begin{array}{l l l} 
X^{(1)}(\s)~~ & \text{for} & 0 \le \s \le \frac{\pi}{2}, \\
X^{(2)}(\s) ~~ & \text{for} &  \frac{\pi}{2} \le \s \le \pi. 
\end{array} \right. \label{star}
\eeq 
\end{subequations} 
The star product is associative and the string field action is invariant under the BRST gauge transformation
\beq \label{BRST}
\d \Psi = Q * \e + \Psi * \e - \e * \Psi .  
\eeq 

Now we shall deform the cubic open string field theory on multiple $Dp$-branes in a fashion similar to the
deformation of the cubic open string field theory on multiple space filling $D$-branes \cite{Lee2016i,Lee2017d}. 
Firstly, we extend the range of the world sheet spatial coordinate $\s$ as 
\beq
0 \le \s \le \pi  ~~\Longrightarrow ~~0 \le \s \le 2\pi 
\eeq 
and redefine the star product as 
\begin{subequations}
\beq
\left(\Psi_1 * \Psi_2\right) [X(\s)] &=& \int \prod_{\pi \le \s \le 2\pi} DX^{(1)}(\s) \prod_{0 \le \s \le \pi} DX^{(2)}(\s)  \nn\\
&& 
\prod_{\pi \le \s \le 2\pi} \d \left[X^{(1)}(\s) - X^{(2)}(2\pi -\s) \right] \Psi_1[X^{(1)}(\s)] \Psi_2[X^{(2)}(\s)], \\
X(\s) &=& \left\{ \begin{array}{l l l} 
X^{(1)}(\s)~~ & \text{for} & 0 \le \s \le \pi, \\
X^{(2)}(\s) ~~ & \text{for} &  \pi \le \s \le 2\pi. 
\end{array} \right. \label{star2}
\eeq 
\end{subequations} 
To be consistent, the normal mode expansions of the string coordinates $X^I$, $I=0, 1, \dots, d$ are to be 
also redefined as
\begin{subequations}
\beq
X^\m (\s) &=& x^\m + \sqrt{2} \sum_{n=1} x^\m_n \cos \left(\frac{n}{2} \s\right), ~~~ \m = 0, 1, \dots, p, \label{xmodem2} \\
X^i (\s) &=& \sqrt{2} \sum_{n=1} x^i_n \sin \left(\frac{n}{2} \s\right), ~~~ i = p+1, \dots, d. \label{xmodei2}
\eeq 
\end{subequations} 

\begin{figure}[htbp]
   \begin {center}
    \epsfxsize=0.5\hsize

	\epsfbox{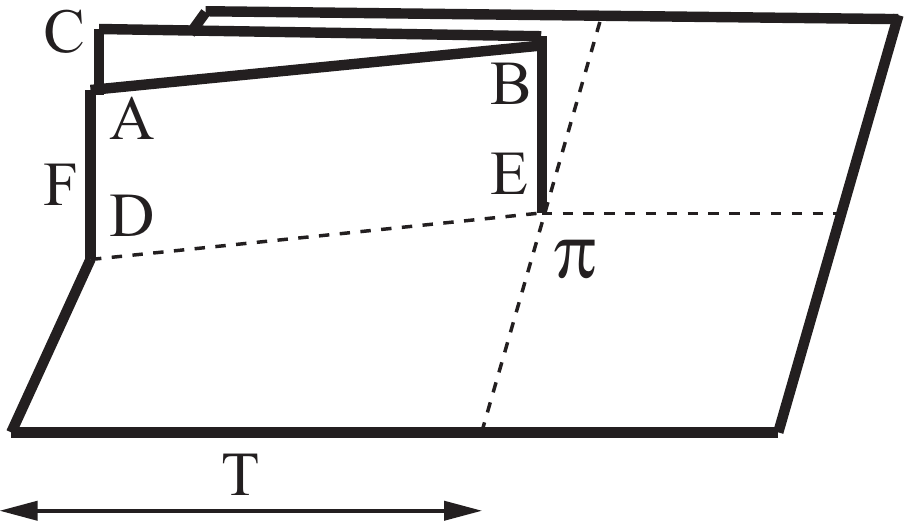}
   \end {center}
   \caption {\label{starpic} The world sheet diagram of the three-string scattering.}
\end{figure}

\begin{figure}[htbp]
   \begin {center}
    \epsfxsize=0.5\hsize

	\epsfbox{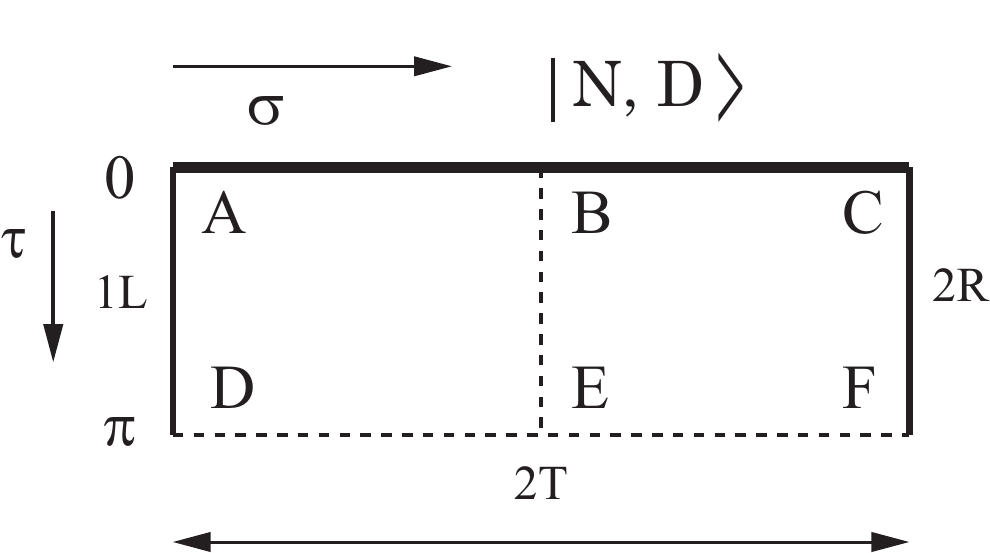}
   \end {center}
   \caption {\label{edge2} Auxiliary patch to be removed effectively by deformation.}
\end{figure}

The Fig. \ref{starpic} depicts the world sheet diagram of three-string scattering. The world sheet of three-string interaction described by the cubic string field theory is not planar but a conic surface with an excess angle $\pi$. 
It is this non-planarity that hinders us from applying the fully developed techniques of the light-cone string field theory to obtain the Fock space representations of multi-string vertices. 
In recent works \cite{Lee2016i,Lee2017d}, we discuss the deformation of the cubic open string field theory on multiple space filling $D$-branes and application of the light-cone string field theory technique to the covariant string field theory. Our discussion on the cubic open string field theory on multiple $Dp$-branes will be parallel to the previous one. 
As we may see in Fig. \ref{starpic}, in the process of three-string scattering physical information,
encoded on the 
half of the first string $\overline{AD}$ and the half of the second string $\overline{CF}$ are not carried 
over to the third string. Thus, the roles of these halves of two strings are auxiliary, and it may be appropriate 
to encode physical information only on the other halves of the two strings. The strings satisfy the Neumann condition or the Dirichlet condition on the boundary $\overline{ABC}$, depending on whether the string coordinate $X^I$ is parallel or perpendicular to the world volume of the $Dp$-branes. It is convenient to separate the
auxiliary patch (Fig. \ref{edge2}) $M_A$ from the rest of the world sheet of the three-string scattering. 
On the patch we may redefine the local coordinates by interchanging the temporal coordinate $\t$ and the spatial
coordinate $\s$, $\t \leftrightarrow \s$. In accordance with the local coordinates we redefine the 
string coordinates $X^I$, $I=0, 1, \dots, d$ as follows 
\beq
X^I (\s) &=& x^I + \sqrt{2} \sum_{n=1} x^I_n \cos \left(\frac{n\pi \s}{2T}\right)\nn\\
&=& x^I + \sum_{n=1} \frac{i}{\sqrt{n}} \left(a^I_n - a^{I\dag}_n \right) \cos \left(\frac{n\pi \s}{2T}\right),
~~~ I= 0, 1, \dots, d, \label{xmodem3}
\eeq  
and express the string state on $\overline{ABC}$ as the following boundary state 
\beq \label{statend}
\vert N, D \rangle = c\, \exp \left( - \frac{1}{2} \sum_{n=1} a^{\m\dag}_n a^{\n\dag}_n \eta_{\m\n} 
+ \frac{1}{2} \sum_{n=1} a^{i\dag}_n a^{j \dag}_n \eta_{ij} \right) \vert 0 \rangle,
\eeq 
satisfying the boundary condition 
\beq
\p_\t X^\m \vert N, D \rangle = 0, ~~~  \p_\s X^i \vert N, D \rangle = 0.
\eeq 

If we choose the Neumann condition as the boundary conditions for the end points of the string on the patch,
we may think of the patch as a world sheet of an open string propagating freely from the initial state 
on $\overline{ABC}$ to the final state on $\overline{DEF}$. Then we find that the string state on $\overline{DEF}$ turns out to be the state $\vert N, D\rangle$ Eq. (\ref{statend}) again 
\beq
\exp\left(-i \pi L_0 \right) \vert N, D \rangle =  \vert N, D \rangle ,
\eeq 
and the Polyakov string path integral over the patch $M_A$ does not contribute to the string scattering amplitude 
because 
\beq
\int_{M_A} \exp(iS) = \langle N, D \vert e^{-i\pi L_0} \vert N, D \rangle = 1. 
\eeq
Therefore, we may effectively remove this auxiliary patch $M_A$ from the non-planar world sheet to render 
the diagram planar. 

It follows from consideration of this deformation that the initial states of the first string and the second string should be given as 
\begin{subequations}
\beq \label{externalstate}
\vert N_1 \rangle \otimes \vert \Psi_1 \rangle , ~~~ \vert \Psi_2 \rangle \otimes \vert N_2 \rangle,
\eeq 
where 
\beq
\vert N_1 \rangle = e^{- \half \sum_{n=1} a^{(1)\dag}_n a^{(1)\dag}_n} \vert 0 \rangle, 
~~~
\vert N_2 \rangle = e^{- \half \sum_{n=1} a^{(2)\dag}_n a^{(2)\dag}_n} \vert 0 \rangle.
\eeq 
\end{subequations}
Here the oscillator operators $a^{(1)\dag}_n$ and $a^{(2)\dag}_n$ act only on the left half of the first string and the right half of the second string respectively. 
As discussed in Refs. \cite{Bordes,Abdu,Rastelli,Gross,Kawano},
we may treat a single string as two halves in string field theory. 
We choose a particular string state to encode the physical information only on the one of halves for 
the first and second strings. 
It would be more convenient to express the external string state  $\vert N_1 \rangle \otimes \vert \Psi_1 \rangle$, Eq. (\ref{externalstate}) in the momentum space. Let us denote the 
string momentum operator on the original (undeformed) string as $\tilde P(\s)$.
\beq
\tilde P(\s) &=& \frac{1}{2\pi} \left\{\tilde p + \sqrt{2} \sum_{n=1} \tilde p_n \cos \left(\frac{n\s}{2}\right)\right\}, ~~~  0 \le \s \le 2\pi .
\eeq 
It may be written also in terms of the string momentum operator defined on the half of the string $P(\s)$ as  
\beq \label{relation}
\tilde P(\s) 
&=& \left\{\begin{array}{ll} 0 & \text{for}~~ \pi <\s \le 2\pi, \\
\frac{1}{\pi} \left(p+ \sqrt{2} \sum_{n=1} p_n \cos(n\s)\right) & \text{for}~~ 0\le \s \le \pi .
\end{array}
\right.
\eeq 
\begin{figure}[htbp]
   \begin {center}
    \epsfxsize=0.3\hsize

	\epsfbox{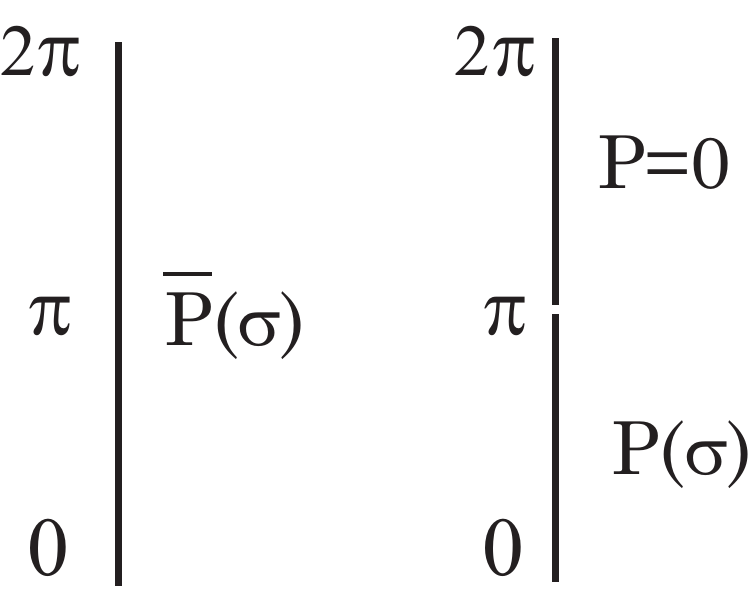}
   \end {center}
   \caption {\label{deformedstate} Comparison of two string momentum bases}
\end{figure}

It is important to note that we deform the cubic open string field theory only by choosing the external string states given as Eq. (\ref{externalstate}) whereas 
the cubic string action is kept intact. Thus, the deformed cubic open string field theory is 
still invariant under the BRST gauge transformation Eq. (\ref{BRST}). 
A simple algebra yields the relation between two momentum operators in terms of normal modes as: 
\beq
\tilde p &=& p, \nn\\
\tilde p_{2k+1} &=& \frac{p}{\pi} \frac{\sqrt{2} (-1)^k}{(k+\half)} + \sum_{n=1}\frac{p_n}{\pi}  \frac{2k(-1)^{k-n}}{k^2-n^2}, ~~~k \ge 0, \\
\tilde p_{2k} &=& p_k , ~~~ k \ge 1. \nn
\eeq 
This relation between two momentum operators implies that the momentum space representations of the 
physical string states $\langle \{n^r_{n}\} \vert \Psi_r \rangle$, $r =1, 2$ are not invariant under the 
deformation. The momentum space 
representations of the physical states transform under the deformation as the momentum space representation
of the number eigen-states $\langle P_r \vert \{n^r_{n}\}\rangle$ change
\beq
\langle P_r \vert \Psi_r \rangle = \int dp^{(r)} \sum_{\{n^r_{n}\}} \langle P_r \vert \{n^r_{n}\} \rangle
\langle \{n^r_{n}\} \vert \Psi_r \rangle, ~~~ r = 1, 2.
\eeq
As we shall show in the paper, if we choose the deformed string states as the external string states, 
we would get the gauge covarint Yang-Mills action directly. We may recall that in the conventional works, 
which make use of the undeformed string state, 
one has to apply the method of field redefinition \cite{David} to the effective string field action to obtain the usual covariant Yang-Mills action. The relation between two momentum operators Eq. (\ref{relation}) 
may allude that deformation of the external string states, adopted in the present work,   
may be equivalent to the procedure of the field redefinition of the conventional works. 

\section{Three String Vertex for Open String on multiple $Dp$-Branes}
 
\begin{figure}[htbp]
   \begin {center}
    \epsfxsize=0.8\hsize

	\epsfbox{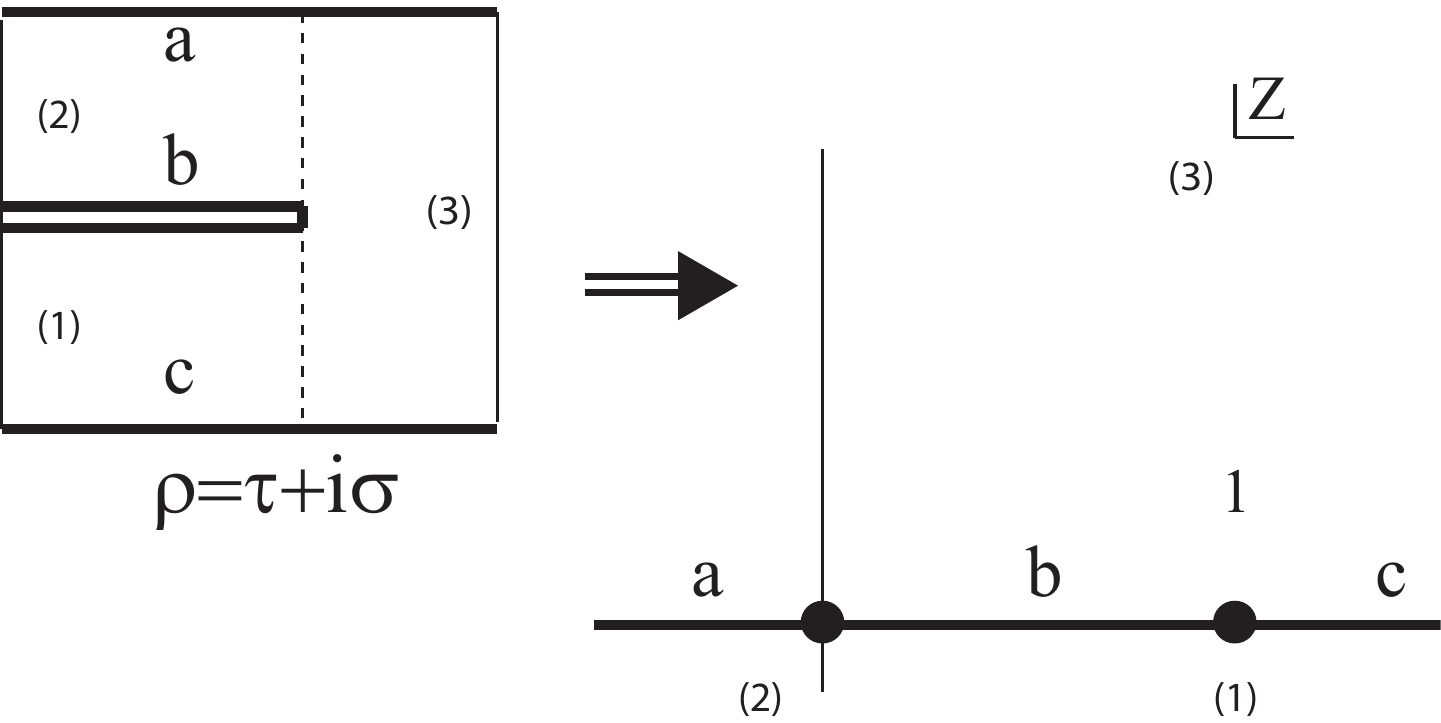}
   \end {center}
   \caption {\label{threestring} Three-String scattering diagram of string field theory 
   in the proper-time gauge.}
\end{figure}

Removing effectively the auxiliary patch from the world sheet diagram of the three-string scattering
by choosing the external string states appropriately, we find that the deformed world sheet diagram is the 
same as the planar diagram of the string field theory in the proper-time gauge \cite{Lee88}: 
It corresponds to the planar world sheet diagram of covaiantized the light-cone string field theory
\cite{Hata1986} with the length parameters which are fixed as
\beq
\a_1 = 1, ~~ \a_2 =1, ~~~ \a_3 =-2.
\eeq 
On the planar world sheet a global coordinate $\rho$ may be introduced such that its real part is the 
proper-time $\text{Re} \rho = \t$ and the planar world sheet may be mapped onto the upper half plane by the 
Schwarz-Christoffel transformation 
\beq
\rho = \sum_r \a_r \ln (z-Z_r) = \ln (z-1) + \ln z, 
\eeq  
where $Z_1=1$, $Z_2=0$, $Z_3 = \infty$. 
The three temporal boundaries labeled as $a$, $b$ and $c$ in Fig. \ref{threestring} are mapped to form
the real line on the upper half plane. The local coordinates on the individual string world sheet patches,
$\z_r$, $r=1, 2, 3$ are related to $z$ as follows: 
\begin{subequations}
\beq
e^{-\zeta_1} &=& e^{\t_0} \frac{1}{z(z-1)}, \\
e^{-\zeta_2} &=& - e^{\t_0} \frac{1}{z(z-1)}, \\
e^{-\zeta_3} &=& - e^{- \frac{\t_0}{2}} \sqrt{z(z-1)} 
\eeq
\end{subequations} 
where $\t_0 = -2 \ln 2$. To obtain the Fock space representtion of the three-string vertex, we need to
solve the Green's equation on the world sheet of the three-string scattering. However, it is not a simple task
to solve the Green's equation directly on the world sheet. The Green's functions on the world sheet may be
obtain by using a comformal transformation (inverse Schwarz-Christoffel transformation) of the well-known the Green's functions on the upper half plane which are given by 
\begin{subequations}
\beq
G_N(z,\zp) &=& \ln \vert z- \zp \vert + \ln \vert z- z^{\prime *} \vert, ~~~\text{for} ~~\text{Neumann boundary condition}, \\
G_D(z,\zp) &=& \ln \vert z- \zp \vert - \ln \vert z- z^{\prime *} \vert, ~~~\text{for} ~~\text{Dirichlet boundary condition}.
\eeq 
\end{subequations} 

Construction of the Fock space representations of multi-string vertices in the case of the Neumann Green's function $G_N$ is well studied in the context of the light-cone string field theory. 
Here we will focus on the construction of 
the Fock space representations by using the Dirichlet Green's function $G_D$. 
We shall begin with the Dirichlet Green's function on an infinite strip (the world sheet of free string propagator).
The strip is mapped onto the upper half plane by a simple conformal transformation 
\beq
\rho = \a \zeta = \a \ln z,
\eeq  
where $\a$ is the length parameter and $\zeta = \xi + i \eta$. The Dirichlet Green's function on the strip is found to be 
\beq
D_{\rm strip}(\zeta, \zeta^\prime)
&=& \ln \vert e^\zeta - e^{\zeta^\prime}\vert - 
\ln \vert e^\zeta - e^{\zeta^{\prime *} }\vert \nn\\
&=& -\sum_{n=1} \frac{2}{n} e^{-n\vert \xi - \xi^\prime\vert} 
\sin n \eta \sin n \eta^\prime .
\eeq 
On the world sheet of multi-string scattering, we may define the Dirichlet functions $\bar D^{rs}_{nm}$, which are 
analogous to the Neumann functions as follows:
\beq \label{dirichlet}
D(\rho_r, \rho^\prime_s) &=& - \d_{rs} \Bigl\{\sum_{n\ge 1} \frac{2}{n} e^{-n\vert \xi_r - \xi^\prime_s \vert}
\sin\left(n\eta_r\right) \sin\left(n\eta^\prime_s\right)
\Bigr\} + 2 \sum_{n, m \ge 0} \bar D^{rs}_{nm} e^{n\xi_r + m\xi^\prime_s} \sin\left(n\eta_r\right) \sin\left(m\eta^\prime_s\right) 
\eeq 
where $\rho_r$ is the coordinate on the patch of the $r$-th string. 
Taking the limit, $\zp \rightarrow Z_s$ or $\zp \rightarrow Z_r$ of Eq. (\ref{dirichlet}), 
we have 
\beq \label{Dn0}
\bar D^{rs}_{n0} = 0, ~~~\text{for}~~ n \ge 0 . 
\eeq
By differentiating Eq. (\ref{dirichlet}) with respect to $\zeta_r$, we find 
\beq 
\bar D^{rs}_{nm} &=& - \frac{1}{nm} \oint_{Z_r} \frac{dz}{2\pi i} \oint_{Z_s} \frac{d z^\prime}{2\pi i} \frac{1}{(z-z^\prime)^2} e^{-n\zeta_r(z) - m \zeta^\prime_s(z^\prime)}, ~~~ n, m \ge 1. 
\eeq 
It turns out that 
\beq\label{Dnm}
\bar D^{rs}_{nm} = - \bar N^{rs}_{nm} .
\eeq 
These results Eq. (\ref{Dn0}) and Eq. (\ref{Dnm}) are not limited to the case of three-string vertex. 
It is interesting that we only need to calculate the Neumann functions to construct the Fock space representations of the multi-string vertices on $Dp$-branes. 

To be explicit, we may write the Fock space 
representation of the three-string vertex in terms of the Neumann function as
\beq 
E[1,2,3] \vert 0 \rangle
&=& \exp \,\Biggl\{ \frac{1}{2} \sum_{r,s =1}^3 \sum_{n, m \ge 1} \bar N^{rs}_{nm} \, 
\a^{(r)\dagger}_{n\m} \a^{(s)\dagger}_{m\n} \eta^{\m\n} + 
\sum_{r=1}^3 \sum_{n \ge 1} \bar N^r_n \a^{(r)\dag}_{n\m} \bbP^\m \nn\\
&& +\t_0 \sum_{r=1}^3 \frac{1}{\a_r} \left(\frac{(p^{(r)}_\m p^{(r)\m}}{2} -1 \right)
- \frac{1}{2} \sum_{r,s =1}^3 \sum_{n, m \ge 1} \bar N^{rs}_{nm} \, 
\a^{(r)\dagger}_{n i} \a^{(s)\dagger}_{m j} \eta^{ij} \Biggr\} \vert 0 \rangle, \label{E3}
\eeq
where $\bbP= p^{(2)}- p^{(1)} $ . 
The three-string interaction may be written as 
\beq \label{3string}
{\cal S}_{[3]} = \int \prod_{r=1}^3 dp^{(r)} \d \left(\sum_{r=1}^3 p^{(r)} \right)  \frac{2g}{3} \langle \Psi_1, \Psi_2, \Psi_3 \vert E[1,2,3] \vert 0 \rangle.
\eeq 

\section{Zero-Slope Limit of the Three-String Interaction}

In the zero-slope limit, the external string states correpond to massless gauge fields $A^\m$ or massless scalar fields $\varphi^i$. By choosing the external string state as follows 
\beq
\langle \Psi^{(1)}, \Psi^{(2)}, \Psi^{(3)} \vert= \Bigl\langle 0 \Bigl\vert \prod_{r=1}^3\Bigl(A_\m(p^{(r)}) a^{(r)}_{1\n} \eta^{\m\n}
+ \varphi_i(p^{(r)}) a^{(r)}_{1j} \eta^{ij} \Bigr ),
\eeq  
we can evaluate the effective interaction between the gauge fields $A^\m$ and the scalar fields $\varphi^i$
which describes the three-string interaction Eq. (\ref{E3}) and Eq. (\ref{3string}) in the zero-slope limit:
\beq \label{s3low}
{\cal S}_{[3]}
&=&  \int \prod_{r=1}^3 dp^{(r)} \d \left(\sum_{r=1}^3 p^{(r)} \right) \frac{2g}{3}\,\text{tr}\,\Bigl\langle 0 \Bigl\vert \prod_{r=1}^3\Bigl\{A_\m(p^{(r)}) a^{(r)}_{1\n} \eta^{\m\n}+ \varphi_i(p^{(r)}) a^{(r)}_{1j} \eta^{ij}\Bigr\}\exp  \left[E[1,2,3] \right] \Bigr\vert 0 \Bigr\rangle \nn\\
&=& \frac{2g}{3} e^{-\t_0 \sum_{r=1}^3 \frac{1}{\a_r}}  \int \prod_{r=1}^3 dp^{(r)} \d \left(\sum_{r=1}^3 p^{(r)} \right) \text{tr}\,\Bigl\langle 0 \Bigl\vert \prod_{r=1}^3\Bigl\{A_\m(p^{(r)}) a^{(r)}_{1\n} \eta^{\m\n}+ \varphi_i(p^{(r)}) a^{(r)}_{1j} \eta^{ij}\Bigr\} \nn\\
&& \left(\frac{1}{2} \sum_{r,s =1}^3 \bar N^{rs}_{11} \, 
a^{(r)\dagger}_{1\m} a^{(s)\dagger}_{1\n}\eta^{\m\n} - \frac{1}{2} \sum_{r,s =1}^3 \bar N^{rs}_{11} \, 
a^{(r)\dagger}_{1 i} a^{(s)\dagger}_{1 j} \eta^{ij}\right) \left(\sum_{r=1}^3 \bar N^r_1 a^{(r)\dag}_1 \cdot \bbP \right) \Bigr\vert 0 \Bigr\rangle .
\eeq
From Eq. (\ref{s3low}) it is clear that we only get a three-gauge interaction term $S_{AAA}$ and an interaction term of type $S_{A\varphi\varphi}$. In the previous works \cite{Lee2016i,Lee2017d} we have evaluated the three-gauge interaction term $S_{AAA}$
\beq \label{SAAA}
S_{AAA} &=& g_{YM} \int \prod_{i=1} dp^{(i)} \d \left(\sum_{i=1}^3 p^{(i)} \right) (p^\m_1 - p^\m_2) \,\text{tr}\, \Bigl( A_\n(p_1)  A^\n(p_2) A(p_3)^\m \Bigr) \nn\\
&=& -g_{YM} \int d^{p+1} x i \, \text{tr}\, \left(\p_\m A_\n - \p_\n A_\m \right) \left[A^\m, A^\n \right]
\eeq
where $g_{YM}$ is the Yang-Mills coupling constant
\beq
g_{YM} = \left(\ap \right)^{\frac{p+1}{4}-1 } g. 
\eeq 

Here we only need to evaluate the term $S_{A\varphi \varphi}$:
\beq \label{SApp}
{\cal S}_{A\varphi\varphi} 
&=& -\frac{2g_{YM}}{3}\times 2^3 \times 2! \int \prod_{r=1}^3 dp^{(r)} \d \left(\sum_{r=1}^3 p^{(r)} \right) \left(p^{\m}_2 - p^{\m}_1 \right) \nn\\
&&\text{tr} \Biggl\{ - \frac{1}{2^4} \varphi_i(p_1) \varphi_j(p_2) \eta^{ij} A_\m(p_3) + \frac{1}{2^3}
\varphi_i(p_2) \varphi_j(p_3) \eta^{ij} A_\m(p_1) + \frac{1}{2^3}\varphi_i(p_3) \varphi_j(p_1) \eta^{ij} A_\m(p_2) \Biggl\} \nn\\
&=& 2 g_{YM} \int \prod_{r=1}^3 dp^{(r)} \d \left(\sum_{r=1}^3 p^{(r)} \right) p^\m_1 \, \text{tr}\, 
\Bigl(\varphi_i(p_1)\left[A_\m(p_3), \varphi^i(p_2) \right]  \Bigr) \nn\\
&=& -2 g_{YM} \int d^d x \,i\, \text{tr} \,  \p_\m \varphi_i \left[ A_\m, \varphi^i \right] . 
\eeq 
Putting two interaction terms together Eq. (\ref{SAAA}) and Eq. (\ref{SApp}), we get the 
cubic interaction term in the zero-slope limit:
\beq
{\cal S}_{[3]} &=& {\cal S}_{AAA} + {\cal S}_{A\varphi\varphi} \nn\\
&=& -ig_{YM} \int d^d x \, \text{tr} \, \Bigl\{\left(\p_\m A_\n - \p_\n A_\m \right) \left[A^\m, A^\n \right] +2 \p_\m \varphi_i \left[ A_\m, \varphi^i \right] \Bigr\}.
\eeq

\section{Zero-Slope Limit of the Four-String Interaction}

The four-string scattering amplitude may be written at the tree level as 
\beq \label{fourscattering}
{\cal F}_{\text{Tree}[4]} &=& \int D[\Psi]\, \text{tr} \prod_{r=1}^4 \Psi^{(r)} \frac{1}{2!} 
\left(\frac{2g}{3}\right)^2 \left[\int \text{tr} \left(\Psi * \Psi* \Psi \right) \right]^2 
e^{\left[-i \int \text{tr} \Psi L_0 \Psi \right]}\, .
\eeq 
The Wick contraction brings us to nine identical Feynman diagrams. 
We may deform the cubic string field theory at two level: 1) We may deform the theory only by choosing 
external string states where the physical information is encoded only on the halves of the external strings. 
2) We may deform the theory at the level of string field action. In the first case where we still keep the 
Witten's cubic string field action, we would get nine Feynman diagrams which are all identical. We only need to take into account of the combinatorics factor as in Eq. (\ref{fourscattering}). In this paper, only the case 1) will be discussed. Of course, we may deform the theory at the level of action also. In the second case we get Feynman diagrams of different types and should worry about the Wick contraction of string field
operators with different length parameters. These problems can be resolved by using the properties of the 
string propagator and the Neumann functions of three-string vertex: The string propagator does not depend
on the length parameters and the Neumann functions of three-string vertex depend only on the ratios
of the length parameters. We would get nine Feynman diagrams of four-string scattering also in this case
which can be made planar. 
Although these Feynman diagrams are not identical, their contributions to the low energy effective action are all idential. The reason is that the string scattering amplitudes which the string Feynman diagrams produce,
only depend on the Koba-Nielsen variables, not on the length parameters. This point has been elaborated in some detail in Ref. \cite{Lee2016i}.

\begin{figure}[htbp]
 \begin {center}
    \epsfxsize=0.7\hsize

	\epsfbox{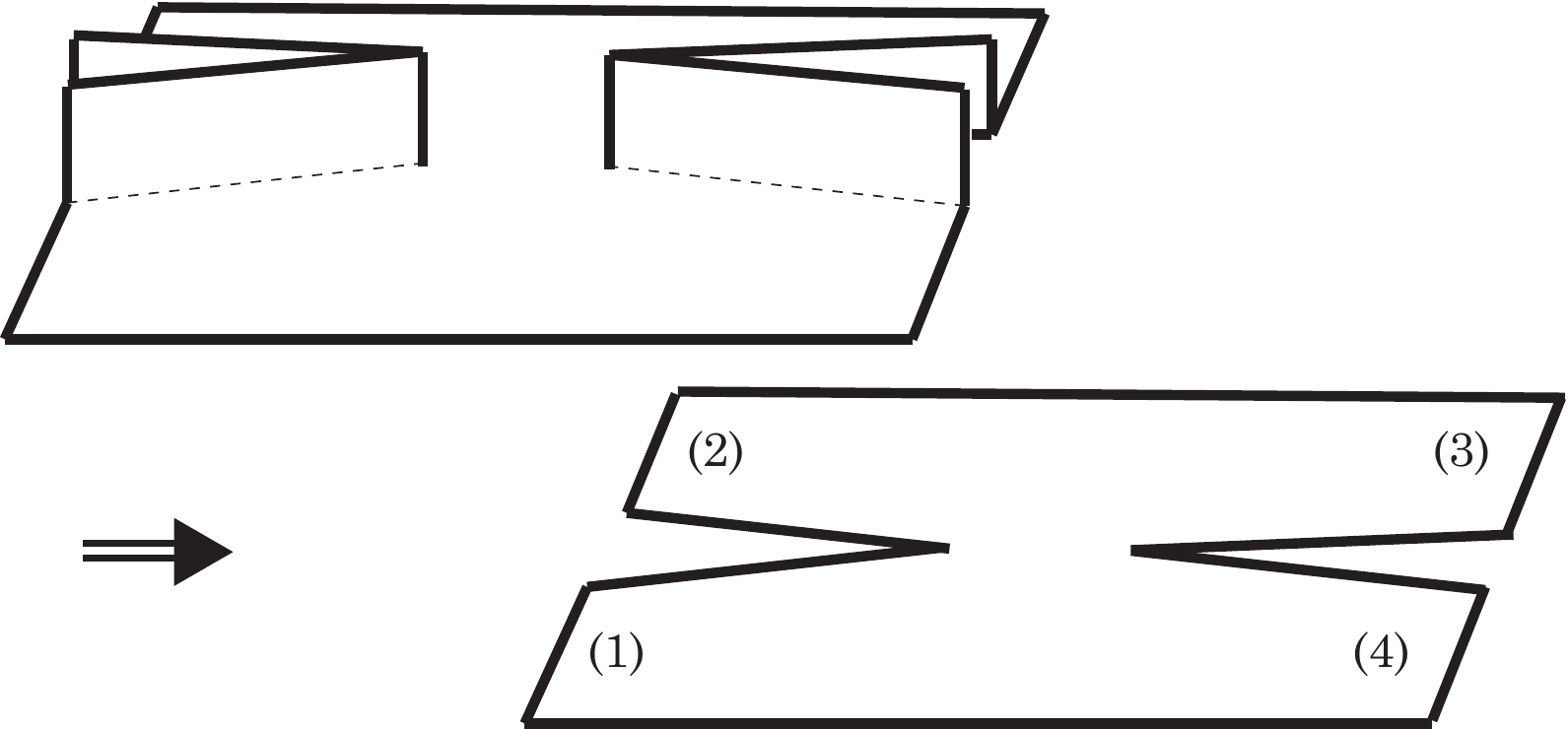}
  \end {center}
   \caption {\label{deformfour2} Deformation of the four-string scattering diagram.}
\end{figure}

If we choose the external string states appropriately to encode physical information only on the halves of the external strings as in the 
case of the three-string scattering, the non-planar diagram of the cubic string field theory may
reduce to the planar diagram of the string field theory in the proper-time gauge as depicted 
in Fig. \ref{deformfour2}. 
Then, by applying the Cremmer-Gervais identity \cite{Cremmer74}, we may cast the four-string scattering amplitude
into a $SL(2,R)$ invariant form:
\begin{subequations} 
\beq 
{\cal F}_{[4]} &=& 2g^2 \int \left\vert\frac{\prod_{r=1}^4 dZ_r }{ dV_{abc}}\right\vert 
\prod_{r<s} \vert Z_r - Z_s \vert^{p_r \cdot p_s} \exp\left[-\sum_{r=1}^4 \bar N^{[4]rr}_{00} \right]\nn\\
&& ~~~~~~\text{tr}\,\bigl\langle \Psi^{(1)}, \Psi^{(2)}, \Psi^{(3)}, \Psi^{(4)} \bigl\vert \exp \left[E_{[4]} \right] \bigr\vert 0 \bigr\rangle , \label{4scattering1}\\
E_{[4]} &=&  \sum_{r, s =1}^4 \left\{ \half \sum_{r,s=1}^4 \sum_{m, n \ge 0} 
\bar N^{[4] rs}_{mn}\a^{(r)\dag}_{m\m} \a^{(s)\dag}_{n\n}\eta^{\m\n}
-\half \sum_{r,s=1}^4 \sum_{m, n =1} \bar N^{[4] rs}_{mn} \a^{(r)\dag}_{m i} \a^{(s)\dag}_{n j}\eta^{ij} 
\right\} . \label{4scattering2}
\eeq
\end{subequations}
The planar diagram Fig. \ref{deformfour2} corresponds to that of light-cone string field theory with length
parameters fixed as 
\beq
\a_1 =1, ~~~ \a_2 =1, ~~~ \a_3 =-1, ~~~ \a_4 =-1 .
\eeq 

We may fix the $SL(2,R)$ invariance by choosing 
\beq
Z_1 =\infty, ~~~ Z_2 =1, ~~~ Z_3 =x, ~~~ Z_4 =0, ~~~ 0 \le x \le 1
\eeq 
where $x$ is the Koba-Nielsen variable of the four-string scattering. 
The Schwarz-Christoffel transformation which maps the four-scattering world sheet onto the upper half plane is 
given as 
\beq
\rho &=& \sum_{r=1}^4 \a_r \ln (z-Z_r)= \ln(z-1) - \ln z - \ln (z-x).
\eeq 
The local coordinates on individual string patches $\zeta_r$, $r=1,2,3,4$ are related to the coordinate 
on the upper half plane $z$ as follows \cite{Lee2017d}:
\begin{subequations}
\beq
e^{-\zeta_1} &=& e^{\t_1} \frac{z(z-x)}{1-z}, ~~~e^{-\zeta_2} = - e^{\t_1} \frac{z(z-x)}{1-z}, \\
e^{-\zeta_3} &=&  e^{-\t_2} \frac{(1-z)}{(z-x) z}, ~~~ e^{-\zeta_4} = - e^{-\t_2} \frac{(1-z)}{(z-x)z} 
\eeq
\end{subequations}
where $\t_1$ and $\t_2$ are two interaction times on the world sheet
\begin{subequations} 
\beq
\t^{(1)}_0 &=& \t^{(2)}_0 =\t_1 = -2 \ln \left(1+ \sqrt{1-x} \right) < 0, \\
\t^{(3)}_0 &=& \t^{(4)}_0 = \t_2 = -2 \ln \left(1- \sqrt{1-x} \right) > 0 . 
\eeq
\end{subequations}

To evaluate the effective action in the zero-slope, limit we choose the external string states as 
\beq \label{4states}
\Bigl\langle \Psi^{(1)}, \Psi^{(2)}, \Psi^{(3)}, \Psi^{(4)} \Bigl\vert  = \Bigl\langle 0 \Bigl\vert \prod_{r=1}^4\Bigl\{A_\m(p^{(r)}) a^{(r)}_{1\n} \eta^{\m\n}
+ \varphi_i(p^{(r)}) a^{(r)}_{1j} \eta^{ij} \Bigr\}.
\eeq  
It is expected from Eqs. (\ref{4scattering1},\ref{4scattering2}) and Eq. (\ref{4states}) that we would obtain interaction terms of following three types:
\beq
AAAA, ~~~ AA\varphi\varphi,  ~~~ \varphi\varphi\varphi\varphi . \nn
\eeq 
In the previous works \cite{Lee2016i,Lee2017d}, we calculated the effective four-gauge field action. The effective four-gauge field 
action, $S^{\text{effective}}_{AAAA}$ obtained by evaluating the four-string scattering amplitude contains 
both the contact quartic gauge field action $S_{AAAA}$ and the effective four-gauge field interaction mediated by massless gauge field $S^{\text{massless}}_{AAAA}$:
\begin{subequations} 
\beq
S^{\text{effective}}_{AAAA} &=& S_{AAAA} + S^{\text{massless}}_{AAAA}, \\
S_{AAAA} &=& \frac{g_{YM}^2}{2} \int d^{p+1} x \, \text{tr} \, 
\left[A^\m, A^\n \right]\left[A_\m, A_\n \right], \label{SAAAA1} \\
S^{\text{massless}}_{AAAA} &=&  g_{YM}^2 \int \prod^4_{i=1} dp^{(i)} \d \left(\sum_{i=1}^4 p^{(i)} \right) \,
\left(1 + \frac{2u}{s} \right) \nn\\
&& \text{tr}\, \Bigl(A_\m(p^{(1)}) A^\m(p^{(2)})A_\n(p^{(3)}) A^\n(p^{(4)}) \Bigr) .
\eeq 
\end{subequations}   
 
\begin{figure}[htbp]
 \begin {center}
    \epsfxsize=0.7\hsize

	\epsfbox{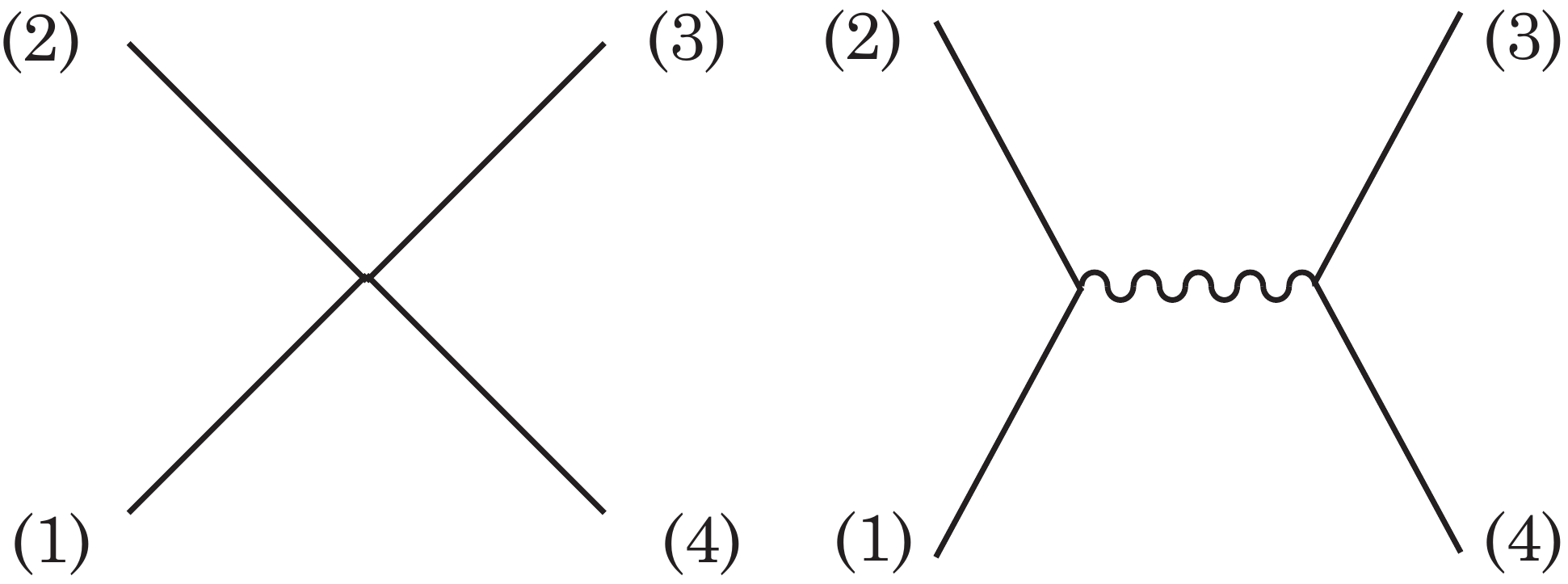}
  \end {center}
   \caption {\label{pppp} Effective four-scalar field interactions.}
\end{figure}

The effective four-scalar field action can be also calculated in a similar way. The four-scalar vertex 
is obtained by choosing the external string states as 
\beq \label{statepppp}
\langle \Psi^{(1)}, \Psi^{(2)}, \Psi^{(3)}, \Psi^{(4)} \vert =  \Bigl\langle 0 \Bigr\vert 
\left\{\prod_{r=1}^4 \varphi_i(p^{(r)}) a^{(r)}_{1j} \eta^{ij} \right\} . 
\eeq 
From Eq. (\ref{4scattering1}) and Eq. (\ref{statepppp}) we find 
\beq
S_{\varphi\varphi\varphi\varphi}^{\text{effective}} &=& g^2_{YM} \int \prod_{r=1}^4 dp^{(r)} \d \left(\sum_{r=1}^4 p^{(r)} \right) 
\int \left\vert\frac{\prod_{r=1}^4 dZ_r }{ dV_{abc}}\right\vert 
\prod_{r<s} \vert Z_r - Z_s \vert^{p_r \cdot p_s} \exp\left[-\sum_{r=1}^4 \bar N^{[4]rr}_{00} \right] \nn\\
&& \text{tr} \Bigl\langle 0 \Bigr\vert 
\left\{\prod_{r=1}^4 \varphi_i(p^{(r)}) a^{(r)}_{1j} \eta^{ij} \right\} \frac{1}{2!} \times \frac{1}{2^2}
\left\{- \sum_{r, s =1}^4 \bar N^{[4] rs}_{mn} a^{(r)\dag}_{1 i} a^{(s)\dag}_{1 j}\eta^{ij} \right\}^2
\Bigr\vert 0 \Bigr\rangle .
\eeq 
The four-scalar field action may be calculated as 
\beq \label{fourscalar}
S_{\varphi\varphi\varphi\varphi}^{\text{effective}} &=& g_{YM}^2\int \prod_{r=1}^4 dp^{(r)} \d \left(\sum_{r=1}^4 p^{(r)} \right) \int_0^1 dx\,\, \text{tr} \Bigl(x^{-\frac{s}{2}} (1-x)^{-\frac{t}{2}} 
\varphi^i(p_1) \varphi^j(p_2) \varphi_i(p_3) \varphi_j(p_4)  + 
\nn\\
&&  + 2 x^{-\frac{s}{2}-2} (1-x)^{-\frac{t}{2}}  \varphi(p_1)^i \varphi(p_2)_i \varphi(p_3)^j \varphi(p_4)_j\Bigr) \nn\\
&=& g_{YM}^2\int \prod_{r=1}^4 dp^{(r)} \d \left(\sum_{r=1}^4 p^{(r)} \right) \, \text{tr} \, 
\Biggl(\varphi^i(p_1)  \varphi^j(p_2) \varphi_i(p_3) \varphi_j(p_4) \nn\\
&&+ \frac{2u}{s} \varphi^i(p_1)  \varphi_i(p_2) \varphi^j(p_3) \varphi_j(p_4) \Biggr).
\eeq 
Here we define the Mandelstam variables as 
\beq
s = -(p_1+p_2)^2, ~~~ t = -(p_1+p_4)^2, ~~~ u = -(p_1+p_3)^2  
\eeq
and make use of 
\beq 
\int^1_0 dx x^{-\frac{s}{2}} (1-x)^{-\frac{t}{2}} = 1, ~~~~
\int^1_0 dx x^{-\frac{s}{2}-2} (1-x)^{-\frac{t}{2}} = \frac{u}{s} 
\eeq
in the zero-slope limit. 
This effective four-scalar field action $S_{\varphi\varphi\varphi\varphi}^{\text{effective}}$ contain the contact quartic scalar action $S_{\varphi\varphi\varphi\varphi}$ as well as the effective four-scalar field interaction 
induced by intermediate massless gauge field $S_{\varphi\varphi\varphi\varphi}^{\text{massless}}$ as depicted by Fig. \ref{pppp} 

In the zero-slope limit, we have shown that there is an interaction term for scalar fields and the gauge fields
$S_{A\varphi\varphi}$ Eq. (\ref{SApp}). 
This interaction term generates the effective four-scalar field interaction perturbatively which is mediated by the massless gauge field. By making use of the usual Feynmman diagrams, we calculate the effective 
four-scalar field interaction term in the zero-slope limit as 
\beq \label{masslesspppp}
S_{\varphi\varphi\varphi\varphi}^{\text{massless}} &=& -\frac{1}{2!} \times (2!)\, g^2_{YM} \int \prod^4_{r=1} dp^{(r)} \d \left(\sum_{r=1}^4 p^{(r)} \right) \, \text{tr}\, \Bigl(\varphi_i(p^{(1)}) \varphi_j(p^{(2)})\eta^{ij}\left(p^{(1)}_\m- p^{(2)}_\m\right) \nn\\
&& \frac{\eta^{\m\n}}{\left(p^{(1)} + p^{(2)}\right)^2} \left(p^{(3)}_\n- p^{(4)}_\n \right) 
\varphi_k(p^{(3)}) \varphi_l(p^{(4)})\eta^{kl} \Bigr)\nn\\
&=& g_{YM}^2 \int \prod^4_{r=1} dp^{(r)} \d \left(\sum_{i=r}^4 p^{(r)} \right) \,
\left(1 + \frac{2u}{s} \right) \text{tr}\, \Bigl(\varphi_i(p^{(1)}) \varphi_j(p^{(2)})\eta^{ij}\varphi_k(p^{(3)}) \varphi_l(p^{(4)}) \eta^{kl}\Bigr). 
\eeq
From Eq. (\ref{fourscalar}) and Eq. (\ref{masslesspppp}) we may identify 
the contact quartic scalar field action $S_{\varphi\varphi\varphi\varphi}$ : 
\begin{subequations} 
\beq
S_{\varphi\varphi\varphi\varphi}^{\text{effective}} &=& S_{\varphi\varphi\varphi\varphi} + S_{\varphi\varphi\varphi\varphi}^{\text{massless}}, \\
S_{\varphi\varphi\varphi\varphi} &=& g_{YM}^2\int \prod_{r=1}^4 dp^{(r)} \d \left(\sum_{r=1}^4 p^{(r)} \right) \nn\\
&&  \text{tr} \, 
\Biggl(\varphi^i(p^{(1)})  \varphi^j(p^{(2)}) \varphi_i(p^{(3)}) \varphi_j(p^{(4)}) - \varphi^i(p^{(1)}) \varphi_i(p^{(2)})\varphi^j(p^{(3)}) \varphi_j(p^{(4)})\Bigr)\nn\\
&=& \frac{g^2_{YM}}{2} \int d^{p+1} x \, \text{tr}\, \left[\varphi^i, \varphi^j \right] \left[\varphi_i, \varphi_j \right] \label{Spppp1}
\eeq 
\end{subequations}

Now we shall calculate the effective interaction term for the scalar field and the gauge field $S_{AA\varphi\varphi}^{\text{effective}}$ by choosing the external string state as 
\begin{subequations}
\beq \label{stateaapp}
\langle AA\varphi\varphi\vert &=&  \bigl\langle 0 \bigr\vert \Bigl\{\bbA(1)\bbA(2)
\bolvphi(3) \bolvphi(4) + \bbA(1)\bolvphi(2)\bbA(3) \bolvphi(4)
\nn\\
&&+ \bbA(1)\bolvphi(2)\bolvphi(3)\bbA(4) + \bolvphi(1)\bbA(2) \bbA(3)\bolvphi(4) \nn\\
&&+  \bolvphi(1)\bbA(2) \bolvphi(3)\bbA(4) + \bolvphi(1) \bolvphi(2) \bbA(3)\bbA(4) \Bigr\}, 
\eeq 
where 
\beq
\bbA(r) &=& A_\m(p^{(r)}) a^{(r)\m}_{1}, ~~~  \bolvphi(r) = \varphi_i(p^{(r)}) a^{(r)i}_{1}, ~~~
r= 1, 2, 3, 4.
\eeq 
\end{subequations}
Making use of Eq. (\ref{4scattering1}) and Eq. (\ref{stateaapp}) we find 
\beq 
S_{AA\varphi\varphi}^{\text{effective}} &=& - \frac{1}{2!} \times 2 \times 
\frac{1}{2^2} g^2_{YM} \int \prod_{r=1}^4 dp^{(r)} \d \left(\sum_{r=1}^4 p^{(r)} \right) 
\int \left\vert\frac{\prod_{r=1}^4 dZ_r }{ dV_{abc}}\right\vert 
\prod_{r<s} \vert Z_r - Z_s \vert^{p_r \cdot p_s}  \nn\\
&& \exp\left[-\sum_{r=1}^4 \bar N^{[4]rr}_{00} \right]\text{tr} \,  \Bigl\langle AA\varphi \varphi \Bigr\vert  \Biggl\{\sum_{r,s =1}^4 \bar N^{[4]rs}_{11} \, a^{(r)\dagger}_{1\m} a^{(s)\dagger}_{1\n}\eta^{\m\n}\Biggr\} \nn\\
&& \Biggl\{\sum_{r,s =1}^4 \bar N^{[4]rs}_{11} \, a^{(r)\dagger}_{1 i} a^{(s)\dagger}_{1 j} \eta^{ij}\Biggr\} \Bigr\vert 0 \Bigr\rangle .
\eeq
In terms of the Koba-Nielson variable $x$, we may rewrite $S_{AA\varphi\varphi}^{\text{effective}}$ as 
\beq
S_{AA\varphi\varphi}^{\text{effective}} 
&=& - \frac{g^2_{YM}}{4} \int \prod_{r=1}^4 dp^{(r)} \d \left(\sum_{r=1}^4 p^{(r)} \right) \int_0^1 dx \, x^{-\frac{s}{2}} (1-x)^{-\frac{t}{2}} \nn\\
&& \text{tr}\, \Biggl\{ \frac{1}{x^2} A_\m(p^{(1)}) A^\m(p^{(2)}) \varphi_i(p^{(3)}) \varphi^i (p^{(4)}) + 
A_\m(p^{(1)}) \varphi_i(p^{(2)}) A^\m(p^{(3)})\varphi^i (p^{(4)}) \nn\\
&& + \frac{1}{(1-x)^2} A_\m(p^{(1)})\varphi_i(p^{(2)}) \varphi^i (p^{(3)}) A^\m(p^{(4)}) + 
\frac{1}{(1-x)^2} \varphi_i(p^{(1)}) A_\m(p^{(2)}) A^\m(p^{(3)})\varphi^i (p^{(4)}) \nn\\
&& + \varphi_i(p^{(1)}) A_\m(p^{(2)}) \varphi^i (p^{(3)})A^\m(p^{(4)})+ \frac{1}{x^2} \varphi_i(p^{(1)})  \varphi^i (p^{(2)}) A_\m(p^{(3)}) A^\m(p^{(4)}) \Biggr\} \nn\\
&=& - \frac{g^2_{YM}}{4} \int \prod_{r=1}^4 dp^{(r)} \d \left(\sum_{r=1}^4 p^{(r)} \right) \Biggl\{ 
\frac{u}{s}  A_\m(p^{(1)}) A^\m(p^{(2)}) \varphi_i(p^{(3)}) \varphi^i (p^{(4)}) \nn\\
&& + A_\m(p^{(1)}) \varphi_i(p^{(2)}) A^\m(p^{(3)})\varphi^i (p^{(4)}) + \frac{u}{t} A_\m(p^{(1)})\varphi_i(p^{(2)}) \varphi^i (p^{(3)}) A^\m(p^{(4)}) \nn\\
&& + \frac{u}{t}\varphi_i(p^{(1)}) A_\m(p^{(2)}) A^\m(p^{(3)})\varphi^i (p^{(4)}) + \varphi_i(p^{(1)}) A_\m(p^{(2)}) \varphi^i (p^{(3)})A^\m(p^{(4)}) \nn\\
&& + \frac{u}{s} \varphi_i(p^{(1)})  \varphi^i (p^{(2)}) A_\m(p^{(3)}) A^\m(p^{(4)})
\Biggr\}. \label{effaapp}
\eeq 
By rearranging terms in Eq. (\ref{effaapp}), we may express the effective action in the zero-slope limit  as 
\beq \label{effSAA}
S_{AA\varphi\varphi}^{\text{effective}} &=& - \frac{g^2_{YM}}{2} \int \prod_{r=1}^4 dp^{(r)} \d \left(\sum_{r=1}^4 p^{(r)} \right) \Biggl\{A_\m(p^{(1)}) \varphi_i(p^{(2)}) A^\m(p^{(3)})\varphi^i (p^{(4)}) \nn\\ && ~~~~~~~~~~~ + \frac{2u}{s} A_\m(p^{(1)}) A^\m(p^{(2)}) \varphi_i(p^{(3)}) \varphi^i (p^{(4)}) 
\Biggr\} . 
\eeq

\begin{figure}[htbp]
   \begin {center}
    \epsfxsize=0.6\hsize

	\epsfbox{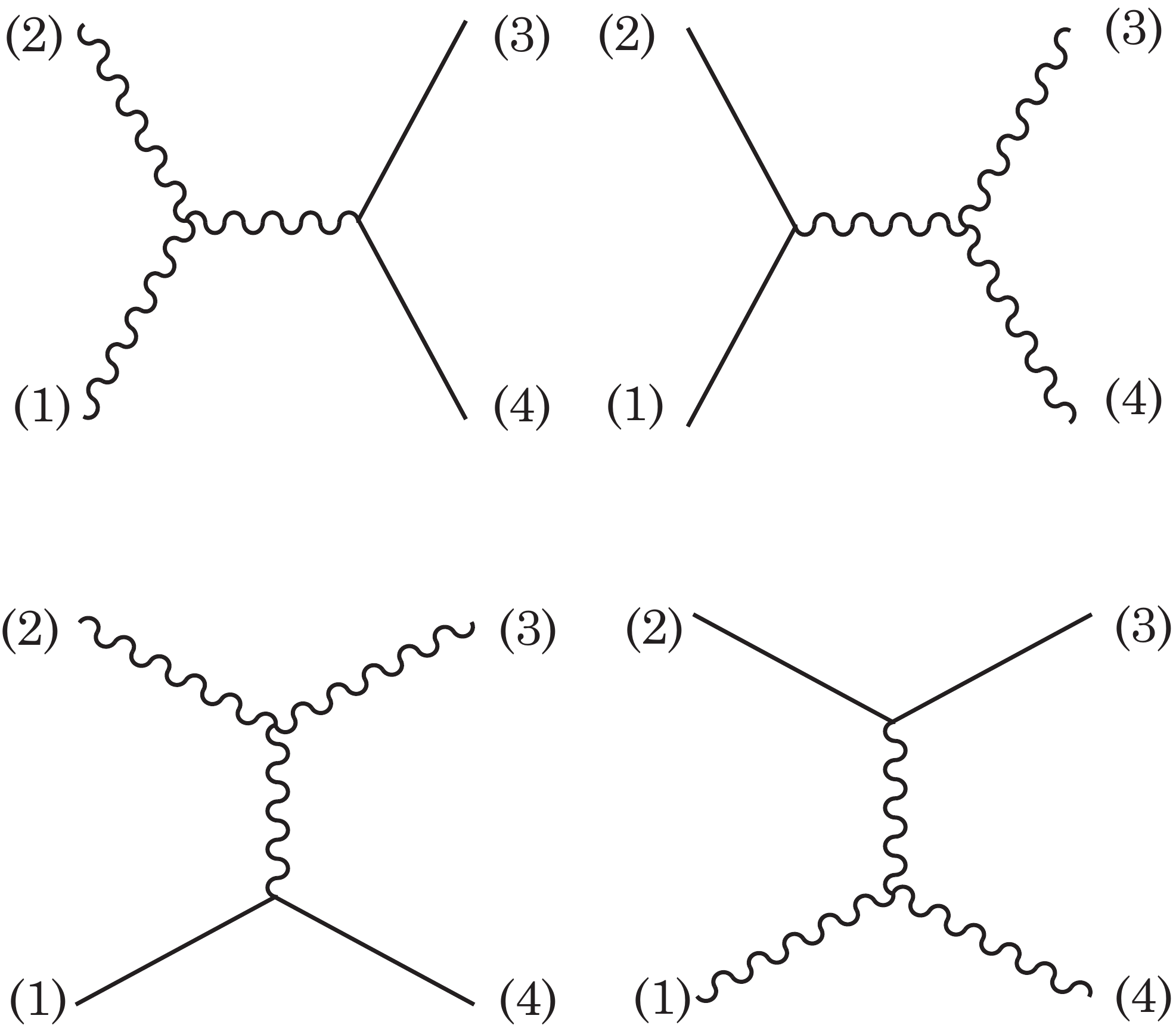}
   \end {center}
   \caption {\label{aapp} Effective gauge-scalar field interactions}
\end{figure}

From $S_{AA\varphi\varphi}^{\text{effective}}$ we should substract the effective gauge-scalar field interaction $S_{AA\varphi\varphi}^{\text{massless}}$ which is generated by the 
cubic intractions, $S_{AAA}$ Eq. (\ref{SAAA}) and $S_{A\varphi\varphi}$ Eq. (\ref{SApp}) to identify the contact 
gauge-scalar field interaction:
\beq 
S_{AA\varphi\varphi}^{\text{effective}} = S_{AA\varphi\varphi}+ S_{AA\varphi\varphi}^{\text{massless}}.
\eeq
The Feynman diagrams corresponding to the effective gauge-scalar field interaction $S_{AA\varphi\varphi}^{\text{massless}}$ which is mediated by massless gauge fields are depicted in 
Fig. \ref{aapp}. The effective gauge-scalar field interaction $S_{AA\varphi\varphi}^{\text{massless}}$
may be evaluated as 
\beq \label{masslessAA}
S_{AA\varphi\varphi}^{\text{massless}} &=& g^2_{YM}  \int \prod_{r=1}^4 dp^{(r)} \d \left(\sum_{r=1}^4 p^{(r)} \right) \, \text{tr} \, \Biggl\{A_\m(p^{(1)}) A^\m(p^{(2)}) \nn\\
&&  \left(p^{(1)}_\rho - p^{(2)}_\rho \right)\frac{\eta^{\rho\s}}{\left(p^{(1)}+ p^{(2)}\right)^2}  
\left(p^{(3)}_\s - p^{(4)}_\s \right) \varphi_i(p_3) \varphi^i (p_4)
\Biggr\} \nn\\
&=& - g^2_{YM}  \int \prod_{r=1}^4 dp^{(r)} \d \left(\sum_{r=1}^4 p^{(r)} \right) \,
\left(1 + \frac{2u}{s} \right)\, \text{tr} \, \Bigl(A_\m(p^{(1)}) A^\m(p^{(2)})\varphi_i(p_3) \varphi^i (p_4)\Bigr). 
\eeq 
If Eq. (\ref{effSAA}) and Eq. (\ref{masslessAA}) are used, the contact interaction term $S_{AA\varphi\varphi}$ is identified as 
\beq \label{AApp2}
S_{AA\varphi\varphi} &=& -g^2_{YM}\, \int \prod_{r=1}^4 dp^{(r)} \d \left(\sum_{r=1}^4 p^{(r)} \right) \,\text{tr}\, \Biggl(A_\m(p^{(1)}) \varphi_i(p^{(2)}) A^\m(p^{(3)})\varphi^i (p^{(4)}) \nn\\
&& -A_\m(p^{(1)}) A^\m(p^{(2)})\varphi_i(p_3) \varphi^i (p_4)\Biggr) \nn\\
&=& - \frac{g^2_{YM}}{2} \, \int d^{p+1} x \, \text{tr}\, \left[A_\m, \varphi_i \right] \left[A^\m, \varphi^i \right] . 
\eeq 
It should be noted that the sign in front of the contact quartic interaction between gauge fields $A_\m$ and scalar fields $\varphi_i$ in Eq. (\ref{AApp2}) differs from those in front of other two contact quartic interactions $S_{AAAA}$ in Eq. (\ref{SAAAA1}) and $S_{\varphi\varphi \varphi \varphi}$ in Eq. (\ref{Spppp1}). 
It plays an important role as we shall see in the next section. If we apply 
a simple dimensional reduction to effective field theory which describes the zero-slope limit of the string 
field theory in the critical dimensions, we would have gotten a different result.

\section{Matrix Models and Cubic String Field Theory in the Zero-Slope Limit}

If we collect the effective actions which are represented by field theoretical actions for the 
$U(N)$ matrix valued gauge fields and scalar fields, we have
\begin{subequations} 
\beq \label{Dpaction}
S &=& S_0 + S_{AAA} + S_{A\varphi\varphi}+ S_{AAAA} + S_{AA\varphi\varphi} + S_{\varphi\varphi\varphi\varphi} \nn\\
&=& \int d^{p+1} x\, \text{tr}\, \Biggl\{ \frac{1}{2} F_{\m\n} F^{\m\n} + \half \left(D_\m \varphi^i \right)^2 + 
\frac{g_{YM}^2}{2}\left[\varphi^i, \varphi^j \right] \left[\varphi_i, \varphi_j \right]  \Biggr\} 
\eeq
where 
\beq
D_\m \varphi^i = \frac{\p \varphi^i}{\p x^\m} - i g_{YM} \left[A_\m, \varphi^i \right].
\eeq 
\end{subequations}
Here $S_0$ is the free field actions for the gauge fields and the scalar fields which may be derived easily 
from the kinetic term of the string free field action $\text{tr}\, \Psi * Q \Psi$.  
It is worthwhile to mention that we obtain the gauge invariant action without redefining fields.
We only need to impose the Lorentz gauge fixing condition; $\p_\m A^\m =0$. 
The resultant action Eq. (\ref{Dpaction}) is precisely the effective field theoretical action on multiple $Dp$-branes which describes 
dynamics of multiple $Dp$-branes in low energy region. 

If $p=0$, the covariant action for the gauge fields will be absent from the action and the
field theoretical action reduces to a quantum mechanical action
\begin{subequations}
\beq \label{D0action}
S = \int dt \, \text{tr} \, \Bigl\{ \half \left(D_t \varphi^i \right)^2 + 
\frac{g_{YM}^2}{2}\left[\varphi^i, \varphi^j \right] \left[\varphi_i, \varphi_j \right] \Bigr\},
\eeq  
where
\beq
D_t \varphi^i = \frac{d \varphi^i}{d t} - i g_{YM} \left[A, \varphi^i \right].
\eeq 
\end{subequations}
This quantum mechanical action Eq. (\ref{D0action}) is the bosonic part of the fundamental action of the BFSS matrix model where the $U(N)$ matrix
valued scalar fields $\varphi^i$, $i=1, \cdots, d$, play the roles of $D0$-brane transverse coordinates. 
We may observe that in Eq. (\ref{D0action}) the gauge field $A$ is auxiliary. However, in deriving the effective action for
$D0$-branes from the string field theory, we should treat the gauge field $A$ as a dynamical one. Otherwise,
we could not get the correct contact four-scalar interaction term.

What is more interesting is the case where $p=-1$:  
If $p=-1$, all string coordinates $X^I$, $I= 0, 1, \dots, d$, satisfy the Dirichlet condition so that there are no zero modes of string coordinates and their 
canonical conjugates. 
The string field action on the multiple 
$D$-instanton may be written as 
\beq
{\cal S} = \text{tr}\, \Biggl\{\langle \Psi \vert (N-1) \vert \Psi \rangle + \frac{2g}{3}  
\langle \Psi \vert \Psi * \Psi \rangle \Biggr\}, 
\eeq 
where $N$ is the total number operator
\beq
N = \sum_{n=1} n\, a^{\dag}_{n I} a_{nJ} \eta^{IJ}.
\eeq   
The free string field action, $\text{tr} \langle \Psi \vert (N-1) \vert \Psi \rangle$ vanishes for the vector field states $ \vert\varphi^I \rangle$, 
$I = 0, 1, \dots, d $. The Fock space representation of three-string vertex for the open strings on $D$-instanton is given by
\beq 
E_D[1,2,3] \vert 0 \rangle
&=& \exp \,\Biggl\{ 
- \frac{1}{2} \sum_{r,s =1}^3 \sum_{n, m \ge 1} \bar N^{rs}_{nm} \, 
\a^{(r)\dagger}_{n I} \a^{(s)\dagger}_{m J} \eta^{IJ} - \t_0 \sum^3_{r=1} \frac{1}{\a_r} \Biggr\} \vert 0 \rangle, \label{E3d}
\eeq
Choosing the external string state as
\beq
\langle \Psi^{(1)}, \Psi^{(2)}, \Psi^{(3)} \vert= \Bigl\langle 0 \Bigl\vert \prod_{r=1}^3\Bigl(\varphi_I a^{(r)}_{1J} \eta^{IJ} \Bigr ),
\eeq  
we find that there is no cubic term for the vector field $\varphi^I$
\beq
S_{\varphi\varphi\varphi} = \frac{2g}{3} \langle \Psi^{(1)}, \Psi^{(2)}, \Psi^{(3)}\vert E[1,2,3]_D \vert 0 \rangle =0.
\eeq

The four-string scattering amplitude for the open strings on multiple $D$-instantons may be expressed as 
\begin{subequations} 
\beq
{\cal F}_{[4]D} &=& 2g^2 \int \left\vert\frac{\prod_{r=1}^4 dZ_r }{ dV_{abc}}\right\vert 
e^{-\sum_{r=1}^4 \bar N_{00}^{rr}} 
~\text{tr}\,\bigl\langle \Psi^{(1)}, \Psi^{(2)}, \Psi^{(3)}, \Psi^{(4)} \bigl\vert \exp \left[E_{[4]D} \right] \bigr\vert 0 \bigr\rangle , \label{4scatteringinstaton1}\\
E_{[4]D} &=&  \sum_{r, s =1}^4 \left\{ 
-\half \sum_{r,s=1}^4 \sum_{m, n =1} \bar N^{[4] rs}_{mn} \a^{(r)\dag}_{m I} \a^{(s)\dag}_{n J}\eta^{IJ} 
\right\} . \label{4scatteringinstanton2}
\eeq
\end{subequations}
We may calculate the effective four-vector interaction by choosing the external string state as 
\beq
\bigl\langle \Psi^{(1)}, \Psi^{(2)}, \Psi^{(3)}, \Psi^{(4)} \bigl\vert =  \Bigl\langle 0 \Bigl\vert \prod_{r=1}^4\Bigl(\varphi_I a^{(r)}_{1J} \eta^{IJ} \Bigr ).
\eeq 
Using Eq. (\ref{4scatteringinstaton1}), we find that the effective four-vector 
interaction is evaluated to be
\beq
S_{\varphi\varphi\varphi\varphi}^{\text{effective}} &=& g^2_{YM}
\int \left\vert\frac{\prod_{r=1}^4 dZ_r }{ dV_{abc}}\right\vert 
e^{-\sum_{r=1}^4 \bar N_{00}^{rr}} 
 \text{tr} \Bigl\langle 0 \Bigr\vert 
\left\{\prod_{r=1}^4 \varphi_I a^{(r)}_{1J} \eta^{IJ} \right\} \nn\\
&&  \frac{1}{2!} \times \frac{1}{2^2}
\left\{- \sum_{r, s =1}^4 \bar N^{[4] rs}_{11} a^{(r)\dag}_{1 I} a^{(s)\dag}_{1 J}\eta^{IJ} \right\}^2
\Bigr\vert 0 \Bigr\rangle .
\eeq 
Making use of the Neumann functions for the four-string vertex in the proper-time gauge \cite{Lee2016i,Lee2017d} leads us to 
\beq \label{fourscalard}
S_{\varphi\varphi\varphi\varphi}^{\text{effective}} &=& g_{YM}^2\, \int_0^1 dx \,\text{tr}\, \Bigl(
\varphi^I \varphi^J \varphi_I \varphi_J 
+ \frac{2}{x^2}\, \varphi^I \varphi_I \varphi^J\varphi_J\Bigr) \nn\\
&=& g^2_{YM} \,\text{tr}\, \Biggl\{ \frac{1}{2} [\varphi^I, \varphi^J ] [\varphi_I, \varphi_J ] + \left(\frac{2}{\e} -1 \right) \left(\varphi^I \varphi_I \right)^2 \Biggr\}.
\eeq
By comparing the effective action with the bosonic part of the IKKT matrix model
\beq
S_{IKKT} &=& \frac{g^2_{YM}}{2}\text{tr}\,  [\varphi^I, \varphi^J ] [ \varphi_I, \varphi_J ], 
\eeq 
we find that the effective matrix action $S_{\varphi\varphi\varphi\varphi}^{\text{effective}}$ 
differs from the bosonic part of the IKKT matrix model action by $\left(\frac{2}{\e} -1 \right)\left(\varphi^I \varphi_I \right)^2$ which is divergent.  
It is hard to conceive that this term arises as an effective interaction between the vector $\varphi^I$ and some other string states at low mass level. We may recall that the vector states $\varphi^I$ are 
only string states at mass level $1$ and there is no cubic interaction term for $\varphi^I$. 
This point may be clarified further in the supersymmetric, BRST invariant string field theory on 
multiple $D$-instantons.

\section{Conclusions and Discussions}

In the present work, we discussed the cubic open string field theories on multiple $Dp$-branes. Interacting 
string field theories have been studied only for the open strings on space filling $D$-branes, which 
satisfy the Neumann boundary conditions. However, it is equally important to explore the 
open string field theory on multiple $Dp$-branes because we can define covariant field theories in dimensions lower than the critical dimensions within the framework of string theory. Even some non-renormalizable quantum field theories may be studied in a consistent manner as  effective theories, describing dynamics of open strings on $Dp$-branes in low energy region. The open string field theories on multiple $D0$-branes and on multiple $D$-instantons are of particular importance as they are intimately related to the matrix model of Banks-Fishler-Shenker-Susskind (BFSS) and the matrix model of Ishibashi-Kawai-Kitazawa-Tsuchiya (IKKT) respectively. 

We define the open string field theory on multiple $Dp$-branes by extending Witten's cubic open string field for string on a space filling $D$-brane \cite{Rastelli2001,Okawa2012}. Then, we have shown that the deformation procedure, developed previously for the Witten's cubic open string field theory, is also applicable to the 
string field theory on multiple $Dp$-branes. By mapping the world sheet for three-string scattering onto
the upper half plane and using the simple Green's functions 
on the upper half plane, we construct the Fock space representation of the three-string vertex.
The Fock space representation of the four-string vertex on multiple $Dp$-branes also has been 
constructed in a similar manner. The effective field theories, describing the open strings on multiple 
$Dp$-branes in the low energy region are obtained by choosing the low mass level string states 
and evaluating the three-string and the four-string scattering amplitudes. The resultant effective 
actions are those of the $U(N)$ matrix valued scalar fields, interacting with $U(N)$ non-Abelian gauge 
fields in $(p+1)$ dimensions. It must be emphasized that we have obtained the gauge covariant effective actions without using the field redefinition or the level truncations in contrast to previous works in the
literature. We only need to impose the usual Lorentz gauge fixing condition. 

If we choose $p=0$, the effective field theory action reduces to a quantum mechanical action, i.e.,
the bosonic part of the fundamental action of the BFSS matrix model. By an explicit evaluation, we confirmed 
it. We also noted that one should treat the gauge field $A$ as a dynamical field when one derive the effective quantum mechanical action from the string field theory on multiple $Dp$-branes, although 
the role of the gauge field becomes auxiliary in the final form of the matrix model action. If we further 
lower $p$ to define the string field theory on multiple $D$-instantons, we expect that the effective theory 
is described by matrices only. However, the effective matrix action contains a divergent term
and differs from the IKKT model action by this divergent term. We could not find a satisfactory resolution for this discrepancy within the framework of the open bosonic string theory. The resolution may be found 
in the complete string theory which is BRST invariant and super-symmetric. 

We have shown that the deformed cubic open string field theory, if properly defined on multiple $Dp$-branes,
correctly captures the dynamics of multiple $Dp$-branes as the theory reduces to the previously known effective gauge covariant field theory. The advantage of the string field theory approach over other 
ones is evident. Because the string field theory possesses full degrees of freedom of the open string, 
it is possible to develop a systematic perturbation theory and to calculate the non-zero slope corrections
which were beyond the scopes of previous approaches. 

Recently, the cubic open string field theories on $Dp$-branes are extended to closed string theory in the proper-time gauge. 
The three-closed-string scattering \cite{TLeeEPJ2018} and the four-closed-string scattering amplutudes 
\cite{TLee2018four}  are calculated. They are shown
to reduce to the three-graviton scattering and the four-graviton scattering amplitudes of the perturbative Einstein gravity
in the zero-slope limit respectively. Because the obtained scattering amplitudes of closed string is valid for the full range of energy, they will be useful to study the ultraviolet completion of qunatum gravity. The open string field theory in the proper-time 
gauge is also found to be useful to study the entanglement entropy of string which may differ from the entanglement 
entropy of quantum field theories: The entanglement entropies of string on $Dp$-branes \cite{TLee2018ent} 
may be finite in contrast to 
their counterparts of quantum field theories. They may help us to understand the black hole entropy as an entanglement 
entropy of string.

\vskip 1cm

\begin{acknowledgments}
This work was supported by Basic Science Research Program through the National Research Foundation of Korea(NRF) funded by the Ministry of Education (2017R1D1A1A02017805).
The author benefited from discussions with participants of IBS (Korea) Strings and Fields Workshop 2017.
\end{acknowledgments}


\end{document}